\documentclass[preprint]{aastex631}
\begin{document}
\title{Impact-induced Vaporization During Accretion of Planetary Bodies}
\author{Adrien Saurety}
\affiliation{Institut de Physique du Globe de Paris\\
1 rue Jussieu\\
Paris, 75005, France}
\author{Razvan Caracas}
\affiliation{Institut de Physique du Globe de Paris\\
1 rue Jussieu\\
Paris, 75005, France}
\affiliation{The Center for Planetary Habitability (PHAB), University of Oslo\\ Oslo, 0371, Norway}
\author{Sean N. Raymond}
\affiliation{Laboratoire d’Astrophysique de Bordeaux, Université de Bordeaux\\ CNRS  B18N\\ Pessac, 33615, France}
\accepted{February 2025}
\submitjournal{ApJL}
\begin{abstract}

Giant impacts dominate the late stages of accretion of rocky planets. They contribute to the heating, melting, and sometimes vaporizing of the bodies involved in the impacts. Due to fractionation during melting and vaporization, planet-building impacts can significantly change the composition and geochemical signatures of rocky objects. Using first-principles molecular dynamics simulations, we analyze the shock behavior of complex realistic silicate systems, representative of both rocky bodies. We introduce a novel criterion for vapor formation that uses entropy calculations to determine the minimum impact velocity required to pass the threshold for vapor production. We derive impact velocity criteria for vapor formation — 7.1 km/s for chondritic bodies — and show that this threshold is reached in 61\% and 89\% of impacts in dynamical simulations of the late stages of accretion with classical and annulus starting configuration (respectively) for analogs of Earth. These outcomes should be nuanced by factors such as the impact angle and the mass of the impacting bodies, which further influence the vaporization dynamics and the resultant material distribution. Our findings indicate that vaporization was common during accretion and likely played a crucial role in shaping the early environments and material properties of terrestrial planets.
\end{abstract}

\vspace{0.5cm}

\section{Introduction}

The impacts are a central process in planet formation.  During the growth of terrestrial planets, some models suggest that the impacts became increasingly energetic as 10-100 km-scale planetesimals accreted into planetary embryos that were $\sim$Mars-sized or larger \citep{kokubo_formation_2000}. Other models favor a first phase of pebble accretion followed by impacts between large planetary embryos at the end of accretion \citep{johansen_pebble_2021}. In any case, the final phase of terrestrial accretion was dominated by giant impacts between planetary embryos \citep{morbidelli_building_2012,raymond_building_2009}, culminating with the Moon-forming impact \citep{canup_origin_2001,cuk_making_2012}.  While this phase has been well studied with N-body simulations \citep{agnor_character_1999,chambers_making_2001,clement_mars_2018,fischer_dynamics_2014,izidoro_terrestrial_2015,morbidelli_building_2012,quintana_effect_2016,raymond_building_2009}, including some studies that include the effect of collisional fragmentation \citep{chambers_late-stage_2013,clement_early_2019,emsenhuber_realistic_2020}, such simulations cannot currently capture the full, complex outcomes of giant impacts \citep{genda_merging_2012,leinhardt_collisions_2012}.

The energies reached during giant impacts could sometimes be sufficiently high to melt or even partially vaporize the impactor and target \citep{davies_silicate_2020}. Vaporization would enhance the loss of volatiles in space due to ionization and may significantly affect the geochemical composition of rocky objects in terms of volatile and moderately volatile elements and light isotopes \citep{norris_earths_2017,hin_magnesium_2017}. Elemental and isotopic fractionation during partial vaporization might explain specific geochemical differences between chondrites and differentiated planets \citep{pringle_rubidium_2017,oneill_collisional_2008}. It is, therefore, of great importance to estimate whether impacts leading to partial vaporization are frequent events during terrestrial accretion or whether they are rare and have a limited effect on the global geochemical evolution of rocky objects.

Shocks in geological materials have been studied extensively for decades. Using shock experiments, ab initio molecular dynamics and thermodynamic models such as ANEOS \citep{melosh_hydrocode_2007} or SESAME \citep{ragan_shock_1982}, the community proposed the shock equation of the state of several minerals and compounds such as forsterite \citep{ahrens_shock_1972,root_principal_2018}, MgO \citep{mccoy_hugoniot_2019}, feldspars \citep{ahrens_shock_1969,asimow_shock_2010}, silica \citep{kraus_shock_2012,melosh_hydrocode_2007} or pure iron \citep{li_partial_2020}. Shocks on complex realistic materials, like dunite \citep{mcqueen_hugoniot_1967}, pyrolite \citep{stewart_shock_2020}, granite \citep{shang_hugoniot_2000}, etc. are less common, and they were often focused on exploring melting relations and the thermodynamic properties of silicate melts \citep{thomas_direct_2013}. The reason is that shock experiments are expensive and time-consuming, providing limited points, while ANEOS and SESAME models rely on strong assumptions. Here, we use ab initio molecular dynamics (AIMD) simulations to study complex silicate's Hugoniot Equation Of State (HEOS). This method has already been applied for feldspar by \cite{kobsch_critical_2020}, forsterite by \cite{root_principal_2018}, and pure iron by \cite{li_partial_2020}. We computed the HEOS and the entropy production during shock for realistic silicate systems representative of planetary bodies: the average composition of a CM2 chondrule (CMc) \citep{friend_composition_2018}, the average composition of a CM2 matrix (CMm) \citep{friend_composition_2018}, and the bulk silicate Earth (pyrolite) \citep{mcdonough_composition_1995}. We obtain an entropy production due to impact that is consistent with the entropy of shocked forsterite obtained experimentally by \citep{davies_silicate_2020}.

\section{Methods}

\subsection{Materials}
We use the Jbilet Winselwan CM2 carbonaceous chondrite \citep{friend_composition_2018} as a model for the chondritic compositions. Carbonaceous chondrites are primitive meteorites that carry material representative of the protoplanetary disk \citep{hewins_chondrules_1997,villeneuve_relationships_2015}. \cite{friend_composition_2018} analyzed the composition of the matrix (CMm) and of several chondrules (CMc) using a microprobe. The chondrules are mostly type I (FeO poor), which are dominant in carbonaceous chondrites\citep{hewins_experimental_2005}. 

The terrestrial composition, or pyrolite, closely follows the bulk silicate Earth model (BSE) \citep{mcdonough_composition_1995} that we used in several previous studies \citep{caracas_no_2023,caracas_thermal_2024}. This composition reflects the bulk terrestrial magma ocean and is a good approximation for the composition of the homogenized protolunar disk. 

All three model compositions used here are realistic representations of solar system bodies. The main relative differences are in the (Mg+Fe)/(Si+Al) ratio, which can influence the degree of polymerization in the melt. Table \ref{tab:compo} lists the three compositions adopted here.

\subsection{{\em Ab initio} simulations}\label{Ap:AI}
We perform ab initio (AI) molecular dynamics  (MD) simulations \citep{car_unified_1985} using the Vienna Ab Initio Simulation Package (VASP) software \citep{kresse_ab_1993,kresse_efficiency_1996,kresse_efficient_1996}. The interatomic forces are computed using the projector-augmented wavefunction (PAW) formalism \citep{blochl_projector_1994} of the density functional theory (DFT)  \citep{hohenberg_inhomogeneous_1964,kohn_self-consistent_1965}. We employ the Perdew-Burke-Ernzerhof functional for the generalized gradient approximation \citep{perdew_rationale_1996}. Atomic cores and core electrons are treated as pseudopotentials. The kinetic energy cut-off of the plane waves outside the PAW spheres is set to 550 eV; the cut-off for the Hänkel functions, which describe the augmented charge inside the PAW spheres, is set to 800 eV. The Brillouin zone is sampled at the Gamma point. The electronic population of the energy states is treated using a Fermi distribution with smearing proportional to the temperature (k$_b$T) \citep{mermin_thermal_1965}.

Simulations are performed in the canonical ensemble with a constant number of particles (N), volume (V), and temperature (T). The temperature is controlled with the Nosé-Hoover thermostat \citep{nose_unified_1984,hoover_canonical_1985}. The time steps are one femtosecond long, the simulations are thermalized for about 1 picosecond, and the production runs between 2 and 10 picoseconds. Those values allow us to have both a good convergence of the pressure, temperature, and energy and enough simulation steps to make good statistics and get representative measures for the thermodynamic properties we are interested in. The timestep of one femtosecond allows the atoms to move slowly enough to have a realistic behavior -too large a timestep will lead the simulation to crash because atomic cores overlap- but fast enough to explore several different configurations in the phase space -too small timestep will increase the simulation time needed to let the atoms move enough to encounter different neighbor. Those values of timestep and simulation time are classic in AIMD simulation and have been used in several studies \citep{caracas_no_2023,caracas_thermal_2024,kobsch_critical_2020,li_partial_2020}.

We use the UMD/magmatomix software \citep{caracas_analyzing_2021} to perform the post-processing of the AIMD simulations. The entropy is computed from the simulation using Thermodynamic Integration methods (see Section \ref{Ap:TI} for more details).

\subsection{Hugoniot Rankine equation equation of states}
Shocks are rapid processes in which materials are subjected to extreme compression and heating during short-impact events. Due to the rapid and intense conditions generated by shocks, these processes are inherently out of equilibrium, presenting significant temperature and pressure gradients that cannot be adequately captured by equilibrium-based equations of state, such as the Vinet or Birch-Murnaghan models, which assume a uniformly equilibrated system. Instead, the Hugoniot-Rankine equations, specifically designed to address these rapid, non-equilibrium changes, provide a more accurate description of the material behavior under such extreme conditions.

The Hugoniot-Rankine equations express the conservation of the energy in the shocked material before and after the shock wave passes \citep{kobsch_critical_2020,forbes_shock_2012,isbell_shock_2005,zeldovich_physics_2002}: 
\begin{equation}
    E-E_0+\frac{1}{2}(P-P_0)(\frac{1}{\rho}-\frac{1}{\rho_0})=0 \label{eq:hg}
\end{equation}
where, E, P, $\rho$ stand for the internal energy, pressure, and density of the material attained during shock, respectively, and the $_0$ index represents the initial state.

The Hugoniot equation of state is sensitive to the initial density and temperature of the material. Consequently, we select three representative densities spanning the range measured by \citep{ostrowski_laboratory_2020} for the chondrule and matrix phase (see table \ref{tab:compo}). For each material, the three densities correspond to the average density reported by \cite{ostrowski_laboratory_2020} ($\rho_{0a}$), the average with the standard deviation added ($\rho_{0h}$), and the average with the standard deviation subtracted ($\rho_{0l}$). This approach allows us to test representative preshock densities for our material. We consider two preimpact temperatures, 1500 K and 10 K (undercooled melt). The higher temperature is realistic of what can be found in a molten body of the protoplanetary disk \citep{boss_convective_2004,gaches_aluminum-26_2020,johansen_harvesting_2018}, and the lower temperature is a proxy for a static simulation of a glassy state. We consider only 3.1 g/cm$^3$ initial density for pyrolite at 300 K as representative of the cold state of the magma ocean.

We perform AIMD simulations along several isotherms to find the density where P, E, and $\rho$ satisfy the Hugoniot Rankine equation. Throughout this work, we call the Hugoniot EOS of a material the curve that verifies the Hugoniot equation at all temperatures and pressures. 

The behavior of a material during shock can also be expressed as a function of particle velocity in the target, $u_p$, and the impactor velocity, $U_p$. In this formulation, the pressure in the target, $P$, is related to $u_p$ \citep{forbes_shock_2012,zeldovich_physics_2002}:
\begin{equation}
    P=a\times u_p^2+b\times u_p + c. \label{eq:p(u)}
\end{equation}

We can obtain a series of $u_p$ values at different states along the Hugoniot EOS from AIMD simulations, using the energy conversion  \citep{isbell_shock_2005}:
\begin{equation}
    u_p = \sqrt{2(E-E_0)}
\end{equation}
which, eventually, allows us to fit the \textit{a}, \textit{b}, and \textit{c} parameters from Eq. \ref{eq:p(u)}. 

Assuming that we are in the planar shock approximation, we can determine the peak pressure \citep{forbes_shock_2012} during the shock by looking for the point where 
\begin{equation}
    P(u_{p})=P(u_{p_{impactor}})
\end{equation}
with 
\begin{equation}    
    u_{p_{impactor}}=U_{impactor}-u_p
\end{equation}

This means that the impactor particles will share their momenta with the impacted particles. Figure \ref{fig:p(u)_plot} shows a real case of determination of the maximum pressure for a CMm target and a CMc impactor, the latter with a speed of 10 km/s.

\subsection{N-body simulations models and classification of planets}\label{sec:nbody}

We analyzed the results of 100 N-body simulations of the late phases of terrestrial accretion (for a discussion of different models of terrestrial planet formation, see \cite{raymond_planet_2022}. Of these, 40 (taken from \cite{raymond_building_2009}) were in the context of the so-called `classical model' of terrestrial planet formation \citep{chambers_making_2001}, and contained a smooth disk of planetesimals and planetary embryos extending from 0.5 au out to close to Jupiter's orbit.  The other 60 simulations (taken from \cite{raymond_empty_2017}) were in the context of the `empty asteroid belt' or `annulus' model  \citep{hansen_formation_2009}, in which planetesimals and planetary embryos were initially distributed in a narrow ring between 0.7 and 1-1.5 au.

Each impact is recorded during those simulations, and the two impacted bodies fuse. At the end of the simulations, only a few bodies are left. Those bodies are the planets formed by the accretion simulation. All bodies remaining at the end of the classical and annulus are shown in figure \ref{fig:nbody-planet}. 

We extracted the bodies with similar masses and orbits to the actual Venus, Earth, and Mars, i.e., bodies in the yellow, blue, or red rectangles in figure \ref{fig:nbody-planet}. For each of them, we extracted the impacts that this planet and all the bodies stuck to the main body have experienced during the accretion. This gives us the complete shock history of the constitutive material of each final planet.

\section{Results}
\subsection{Behavior of the constituent materials during impacts}

During large shocks ({\em i.e.} supersonic), the impacted materials are shocked from their initial state on their Hugoniots. The kinetic energy associated with the shock is largely transformed into entropy.  If the shock is sufficiently energetic, the target and/or impactor can be projected up to molten or even supercritical states \citep{kraus_impact_2015,caracas_no_2023}. After the passage of the shock wave, the thermodynamic trajectories of the bodies involved in the impact are largely quasi-isentropic, as there is no heat exchange with the environment. Depending on the amount of entropy generated during impact, production of vapor may occur during release, when the materials reach the liquid-vapor dome. This makes entropy the most critical parameter in describing the thermodynamics of impacts.

%

\begin{figure}[htbp]
    \centering
    \includegraphics[width=0.95\textwidth]{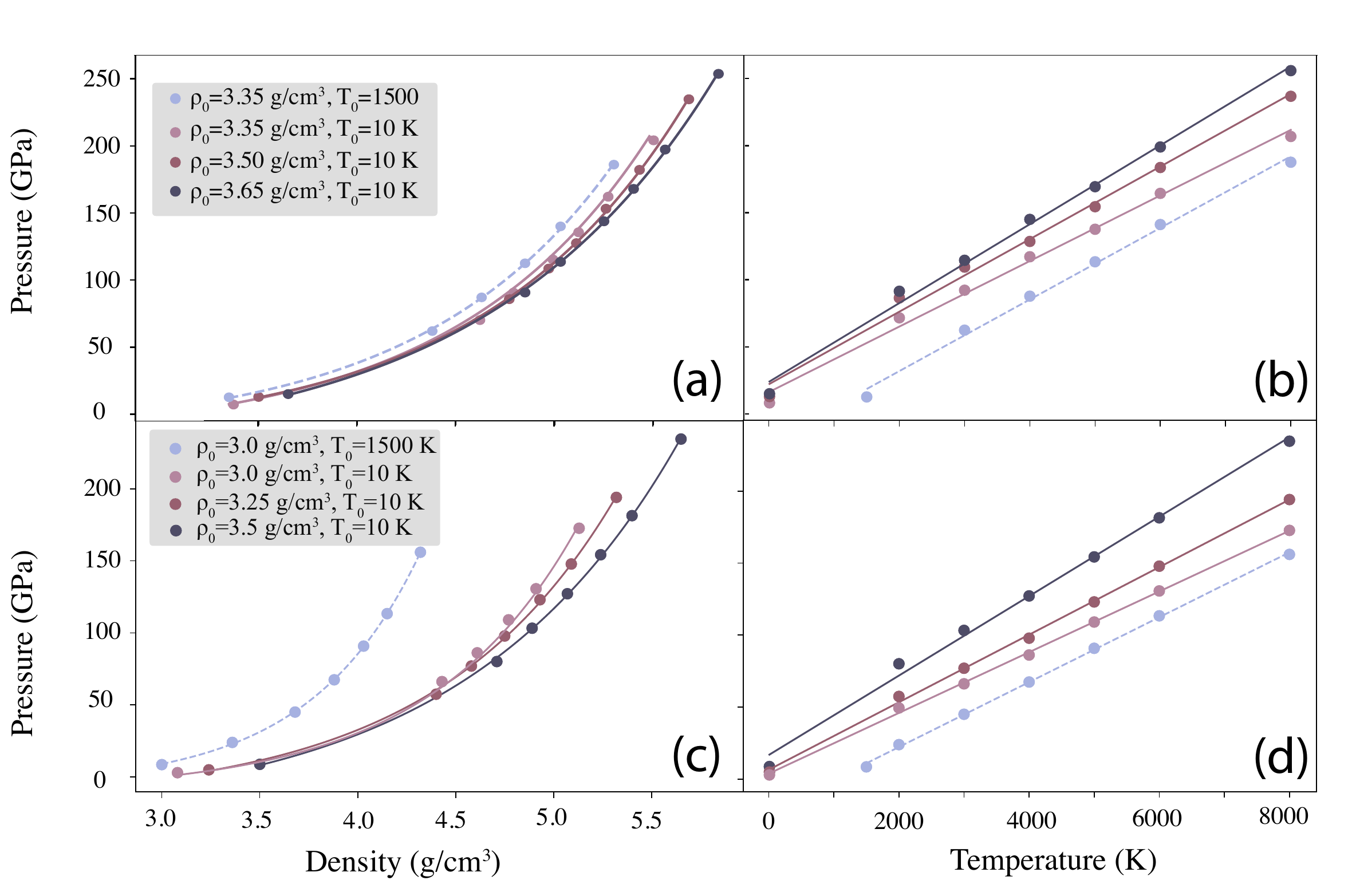}
        \caption{Pressure variation along the Hugoniot shock EOS as a function of density and temperature, for CMc (a and b) and for CMm (c and d) compositions with various initial states. Higher initial densities yield higher peak pressures at a given temperature. Higher initial temperatures yield higher peak temperatures at a given pressure. The effect of the temperature is more important for CMm than for CMc, probably because of larger relative thermal-induced depolymerization of the CMm melt.}
        \label{fig:effect_P_T}
\end{figure}

The topology of the Hugoniot EOS is such that its trajectory is highly dependent on the initial state, \emph{i.e.} $\rho_0$, $T_0$, $E_0$, and $P_0$. To quantify this influence, we independently vary the initial densities and temperatures (see Fig. \ref{fig:effect_P_T}) and compute the Hugoniot EOS for the chondrule composition. We find that for a difference in initial density of 0.3 g/cm$^3$, the EOS diverges by 0.2 g/cm$^3$ by the time the pressure reaches 200 GPa. In terms of temperature, this difference of 0.3 g/cm$^3$ in the initial density leads to a difference of 2000 K at 200 GPa, {\em i.e.} the peak temperature will be 6000 K for a chondrule with $\rho_0=3.65$ g/cm$^3$ and 8000 K for a chondrule with $\rho_0=3.35$ g/cm$^3$. This divergence shows the importance of considering a material with a relevant initial density: the same impact may form a melt and vapor mixture in the first case and a supercritical fluid in the latter. Increasing the temperature from T$_0$=10 K  to T$_0$=1500 K at the same initial density results in an upward shift in density along the entire Hugoniot curve, which is roughly constant: 0.11 g/cm$^3$ at 80 GPa and 0.09 g/cm$^3$ at 180 GPa. In the temperature-pressure phase space, the T$_0$ effect seems less important when the pressure increases and the two Hugoniot curves tend to converge. But it remains non-negligible at moderate pressure with a difference of 1000 K at 150 GPa between the T$_0$=10 K and T$_0$=1500 K curves.

\begin{figure}[htbp]
    \centering
   \includegraphics[width=0.65\textwidth]{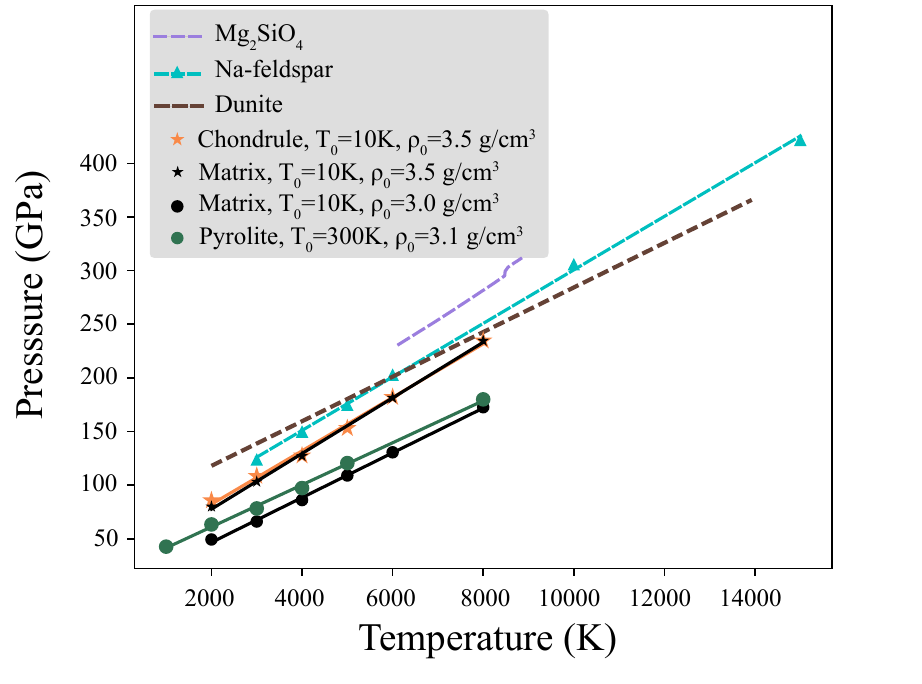}
\caption{Various Hugoniot EOS computed for representative planetary-building silicate compositions (solid lines). All our computed EOS start at low temperatures, either 10 K or 300 K. In the high-temperature and high-pressure regime, the EOS are quasi-linear; this relation is less obvious at low temperatures and low densities. The Hugoniot EOS exhibit similar topologies, but their position is highly dependent on the initial state and, in particular, on the initial density. 
The chondritic and pyrolitic compositions (this study), with a (Mg+Fe)/Si ratio close to mantle values, produce temperatures much higher than the other phases considered here, partly due to their high initial densities.} 
\label{fig:comp_chondru_matrix}
\end{figure}

As $\rho_0$ and T$_0$ clearly have a strong influence on the resulting Hugoniot EOS, we also try to grasp the effect of the chemical composition. We compute the EOS for CMc, CMm, and pyrolite for various initial conditions. For the sake of completeness, we compare the theoretical EOS with the Hugoniot EOS of several representative materials involving planetary impacts obtained from previous experimental or computational studies. Figure \ref{fig:comp_chondru_matrix} shows the differences between all these Hugoniot curves. Although the topologies of the different curves are similar, there is a large scatter in both temperature and pressure. For example, at 150 GPa, the EOS of the Na feldspar, a continental crust mineral, lies at the lowest temperature, only 3000 K, and the EOS of CMm reaches beyond 4500 K. Similarly, at 4000 K, the pressure reached by Na feldspar is 200 GPa, whereas the pressure on CMm is only 130 GPa. Compared to the Hugoniot EOS mentioned above, the ANEOS curve of dunite \citep{root_principal_2018} intersects the EOS of feldspar, but is about 50 GPa above the calculated one of pure forsterite \citep{root_principal_2018}. The two are above CMc, CMm, and pyrolite in terms of pressure. The scatter between the different curves can have multiple causes, some intrinsic, like initial densities, temperatures, degree of polymerization of the materials, role of cations, changes in magnetism, etc., and some extrinsic, like errors in measurements and approximations in calculations. 
Figure \ref{fig:comp_chondru_matrix} also shows that for the same initial conditions of temperature (10 K) and pressure (3.5 g/cm$^3$), the Hugoniot EOS of CMc and CMm are almost overlapping. In contrast, those of pyrolite and CMm for slightly different initial densities are very similar. This suggests that the silicate component of chondritic-like planetesimals can be treated as a homogeneous phase within smooth particle hydrodynamics (SPH) simulations \citep{canup_origin_2001}, representing a considerable simplification in the path to acquire realistic large-scale impact simulations.  

Apart from the initial state, one potential explanation for the scatter and similarities in the pressure and temperature of the different Hugoniot EOS is the difference or similarity in chemistry. The ratio between the number of interstitial cations, such as Na, Ca, Mg, or Fe, and the number of network-forming cations, like Si or Al, induces differences in the degree of silica polymerization between these materials \citep{mysen_structure_1982}. The substantial presence of these structural polyhedra could profoundly influence the dynamic response of the material by strongly structuring the melt. 
\newline

\subsection{Entropy production due to impact}

Passing a shock wave through a material is associated with an increase in entropy. We can link the entropy and the temperature along the Hugoniot using \citep{forbes_shock_2012, zeldovich_physics_2002} equation:
\begin{equation}
    S=A\times ln(T) +B,
\end{equation}
where \textit{A} and \textit{B} are constants. Although we cannot directly compute the entropy and free energy from the AIMD simulations, we can use thermodynamic integration to calculate these numbers. This method provides a direct accurate value for the entropy and does not require indirect measurements. It is, therefore, more reliable than interpolating indirect experimental data on heat capacity, which relies on many approximations \citep{davies_silicate_2020} - or by integrating an ANEOS model, which assumes a set of constant parameters and often gives results significantly different from the experiment \citep{kraus_impact_2015}. Thus, this is the first direct estimate of the entropy produced during a shock and the entropy of a material on the liquid-vapor dome using AIMD simulation and direct entropy measurements.

We calculate the entropy along the Hugoniot EOS of pyrolite and CMc, starting with $\rho_0=3.5$ g/cm$^3$ and $T_0=10$ K. The results are plotted in Figure \ref{fig:entropy}. The entropy of these materials is relatively similar, up to about 6000 K. Below 5000 K, shocks in pyrolite yield higher entropy production than in CMc, and above 5000 K, the tendency is reversed. Under huge shocks, at temperatures far exceeding 6000 K, these two materials show significantly different slopes and tend to diverge. This difference might be explained by the difference between the Hugoniot EOS of CMc and pyrolite (see Figure \ref{fig:comp_chondru_matrix}). Indeed, for the same peak temperature, the peak pressure is higher on the CMc Hugoniot than on the pyrolite Hugoniot. As entropy is a measure of chaos, a significant pressure that slows down the movement of the atoms will also affect the entropy and make it more difficult to increase.

Next, we compute the entropy of the pyrolite at three different temperatures (2000, 4000, and 6000 K) along the liquid spinodal curve from \cite{caracas_no_2023}. Using these values and the critical temperature estimated in \cite{caracas_no_2023}, we can interpolate the shape of the spinodal line in the T - S diagram on the liquid side of the liquid-vapor dome (LVD). The spinodal line marks the limit of metastability: at entropies higher than the spinodal, the pure liquid becomes unstable, and vapor forms. 
\newline

\section{Shock-induced vaporization by impacts during accretion}

\begin{figure}[htbp]
    \centering
   \includegraphics[width=0.65\textwidth]{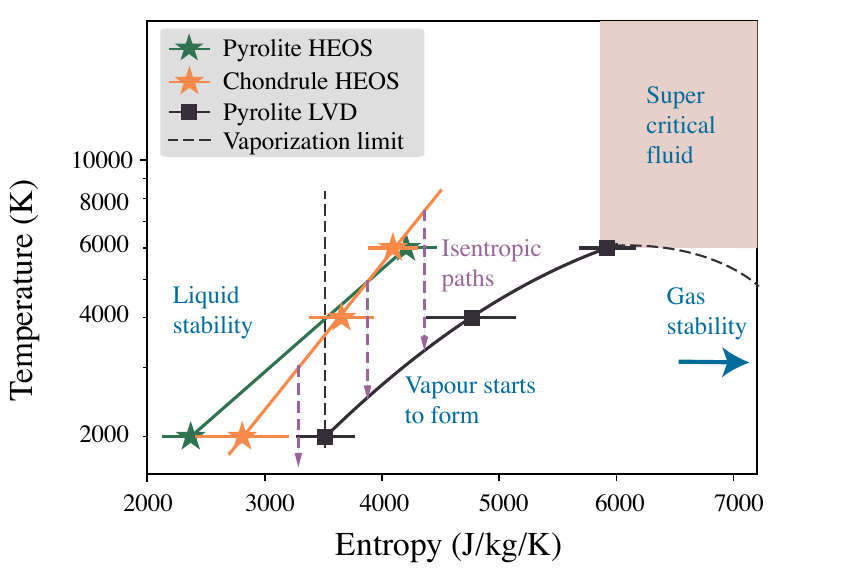}
    \caption{Entropy along the Hugoniot EOS for CMc and pyrolite compositions. The liquid-vapor dome (LVD) is an approximation following the temperature along the liquid spinodal line \citep{caracas_no_2023}. The LVD line is a second-order polynomial fit that ends close to the critical point \citep{caracas_thermal_2024}. The release follows a quasi-isentropic trajectory in the aftermath of an impact, represented by the vertical dashed arrows. Vapor forms if the entropy produced during the shock is larger than the threshold for vaporization (equivalent to the entropy of boiling). Note that the temperature scale is logarithmic.}
    \label{fig:entropy}
\end{figure}

The absolute value of the entropy produced during a shock is useful as a thermodynamic parameter {\em per se}. But its importance becomes evident when considering the post-shock state. In the planar shock approximation, we can link the impact velocity of the impactor to the maximum peak pressure and temperature reached in the impacted body. Then we can estimate the entropy production during shock in the T - S diagram (Figure \ref{fig:entropy}). The shock wave passage causes a rapid increase in pressure and temperature. After the shock wave passes, the material undergoes a decompression. This decompression is a slow process compared to compression and, as such, can be assumed to be an isentropic decompression \citep{forbes_shock_2012,zeldovich_physics_2002,kurosawa_entropy_2013}. Although this assumption has some limitations \citep{heighway_nonisentropic_2019}, it gives a realistic approximation of the conditions encountered during shock release. 

Within the isentropic release assumption, the post-shock evolution of the material in the T - S diagram is a vertical line along which the material cools down at constant entropy. If this vertical line crosses the LVD, partial vaporization occurs. Thus, if the entropy produced during a shock exceeds the entropy of boiling (\textit{i.e.} entropy of incipient vaporization), vapor forms when the system temperature intersects the temperature on the LVD dome. This is our criterion for identifying vapor as a result of an impact. As such, it is a more accurate measure for the incipience of vapor than comparisons to the entropy required to vaporize 50\% of forsterite (as in {\em e.g.} \cite{davies_silicate_2020,kraus_shock_2012}). We then use the liquid-vapor dome calculated by \cite{caracas_no_2023} to obtain the density of molten pyrolite close to the equivalent of a
triple point, at 2000 K, conditions that are similar to the prediction of the ANEOS model for forsterite \citep{davies_silicate_2020}. We calculate the entropy of pyrolite at 2000K close to the 
spinodal line, and find a threshold value of $S_m$=3590~J/kg/K (see Figure \ref{fig:entropy}) to trigger vaporization. This value is close to the triple point entropy for forsterite (3500 J/K/kg) \citep{davies_silicate_2020}.
Our work using a direct measure of the entropy during AIMD simulations gives credence to the results estimated by \cite{davies_silicate_2020} and \cite{kraus_shock_2012}. 

The analysis of the Hugoniot EOS curves (Figure \ref{fig:entropy}) shows that the entropy of boiling at 2000 K is achieved for shock temperatures above $T_{mc}$=3631~K for CMc and above $T_{mp}$=3882~K for pyrolite. These minimal peak temperatures can be related to a minimal peak pressure of $P_{mc}$=122~GPa for CMc and $P_{mp}$=98 GPa for pyrolite (Figure \ref{fig:comp_chondru_matrix}). Finally, we relate the position reached on the Hugoniot EOS curve with the impact velocity ($U_{mc}$), and find that vapor forms for impacts of at least $U_{mc}$ = 7.1 km/s for two chondritic bodies and $U_{mp}$ = 6.1 km/s for two pyrolitic bodies (Figure \ref{fig:up-py}) which is consistent with the 6 km/s critical impact velocity proposed by \cite{davies_silicate_2020} for cold silica. These values are considerably lower than the requirement to vaporize iron cores, at $U_{mc}$ = 11.5 km/s \citep{li_partial_2020}.

We compare this criterion with the impact velocities recorded during dynamical simulations of terrestrial planet accretion, following two different scenarios \citep{raymond_building_2009,raymond_empty_2017}. Figure \ref{fig:histo-all} shows that the annulus model leads to higher impact velocities than the classical model. Higher collision speeds dramatically increase the proportion of impacts that lead to partial vaporization. 

In the approximation of undifferentiated cold silicate bodies, with starting temperatures at 10 K, we find that about 61.5 \% of the impacts are energetic enough to produce vapor when impacting chondritic bodies in the classical accretion simulations. This proportion goes up to 90\% of the impacts in the annulus accretion simulations. A similar analysis for BSE material leads to a proportion of vapor-forming impacts of 68\% and 93\% for classical and annulus cases, respectively. 
These estimations are valid for head-on planar collisions. More realistic impact geometries would reduce these numbers. Taking into account a maximum frequency of impact angles of 45 degrees \citep{pierazzo_understanding_2000}, the threshold velocities are reduced to $U_{mc}$ = 10 km/s and $U_{mp}$ = 8.6 km/s for chondritic and pyrolitic bodies, respectively. This, in turn, reduces the number of impacts that can produce vapor to 43 \% and 51 \% for pyrolite and chondrule -respectively- in the classical model. 

\begin{figure}[htbp]
    \centering
    \includegraphics[width=0.95\textwidth]{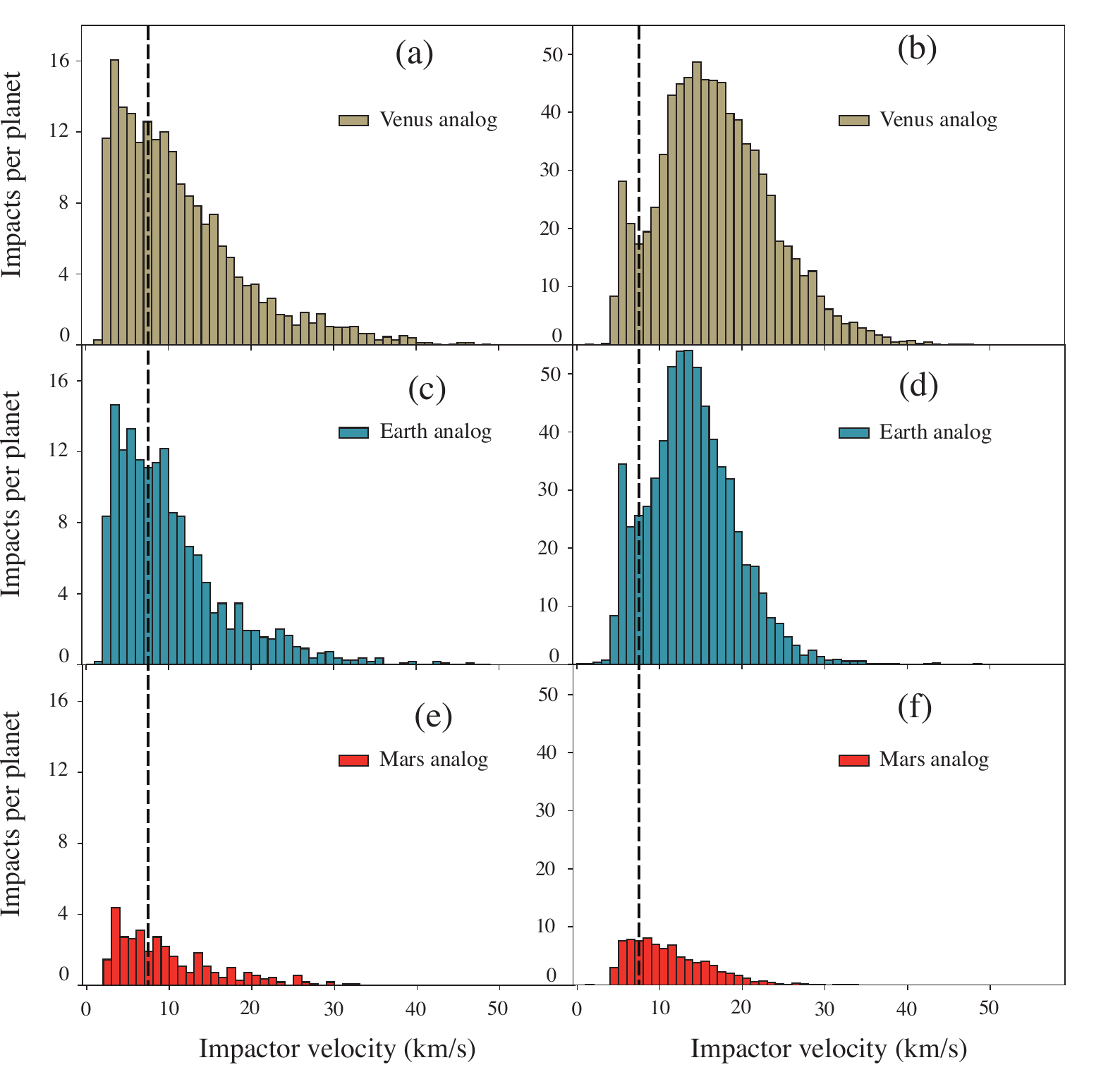}
    \caption{Histogram of the distribution of impact velocities in N-body simulations with a classical starting configuration (panels a, c, and e) and annulus starting configuration (panels b, d, f) for Venus, Earth, and Mars analogs. For the sake of clarity, the vertical scale is not the same on the left and right panels. The dashed lines indicate the critical impactor velocities required to have impact-induced vaporization for chondrule material.}
    \label{fig:histo-comp}
\end{figure}

Furthermore, retracing the sequence of impacts that lead to the formation of Earth analogs (simulated planets with orbits and masses similar to Earth; see Section \ref{sec:nbody}) allows us to evaluate the significance of vapor formation during these events. We find that 61\% and 89\% of the chondritic impacts in the classical and annulus models, respectively, produced vapor (see Figure \ref{fig:histo-comp}). These numbers are comparable to those obtained for Venus analog planets (65\% and 92\% respectively). The proportion of vapor-forming impacts during the history of Mars analogs in the classical model is also close to that of Earth and Venus (56\%). However, Mars experienced less vaporizing impacts in the annulus model than Earth and Venus (77\%). Thus, an analysis of volatile and moderately volatile elements could give credit to the annulus accretion model if it shows that Mars is enriched in volatiles compared to Earth.

A complementary analysis showed that, within the results of N-body simulations, planetesimal impacts occurred in classical and annulus models at higher velocities than embryo impacts. They then present more vaporizing impacts (see Figure \ref{fig:histo-embrypla}). This is due to the higher eccentricity of planetesimals relative to embryos \citep{kokubo_formation_2000}. Moreover, impact velocities do not increase dramatically with time (see Figure \ref{fig:velocity}). It is then unlikely that the dichotomy we see between impact velocities of embryo-embryo and embryo-planetesimal collisions is due to a timing effect. This higher vaporization rate for embryo-planetesimal impacts is consistent with \cite{young_near-equilibrium_2019} and \cite{hin_magnesium_2017}, suggesting that the enrichment of heavy Si and Mg isotopes of differentiated rocky bodies is due to vaporizing events during planetesimal growth.

\section{Conclusion}
Our findings suggest that vapor production during impacts between planetesimals could be a more common phenomenon than previously thought (Figure~\ref{fig:histo-all}). Given the substantial number of vapor-producing impacts, they should be further considered in N-body simulations of impacts. Whether or not vapor is lost post-impact depends significantly on factors such as the mass, velocity, and impact angle of the colliding bodies. Impacts between small planetesimals might produce vapor, but probably only in small amounts, though this could be easily lost by ejection. As discussed above, large impacts would produce more vapor, but this might be trapped in the gravitational potential well that forms after the impact. In this case, the vapor would not escape but fall back into the condensing central body. SPH simulations that specifically use the entropy of the materials involved in the impact, like the ones we computed here, offer advanced predictive capabilities by accurately estimating the volume and fate of vapor produced during impacts. Our findings also suggest that impacts affecting the volatile budget of a planet could occur not only after giant impacts such as the Moon-forming impacts but also may be frequent in smaller-scale impacts. These events, often underestimated, could significantly influence the mass-balance evolution and volatile distribution among accreting planetesimals. This significant proportion of vaporizing impacts we predict could affect our view of the formation of the terrestrial planets. \cite{pringle_rubidium_2017} proposed that the rubidium isotopic signature of the Moon could be due to the loss of rubidium during vaporization events. They suggested that this evaporation could have occurred during the Moon-forming impact or earlier during the accretion of Theia. Potentially different ratios of volatiles and moderately volatile elements in the Martian moons could be related to an impact origin of Phobos. These insights into the dynamics of vaporization during planetary accretion offer pivotal corrections to existing planetary formation and evolution models, potentially altering our understanding of early solar system dynamics.

\section*{Acknowledgement}
AS acknowledges Charles Lelosq's help and constructive comments. RC acknowledges financial support from the UPC Labex UnivEarthS (project VADIS), as well as support from the Research Council of Norway, project numbers 332523 (PHAB), and 325567 (HIDDEN), and access to supercomputing facilities via eDARI stl2816, PRACE RA4947 and RA240046, and Uninet2 NN9697K grant. SNR acknowledges funding from the French Programme National de Planétologie (PNP) and in the framework of the Investments for the Future programme IdEx, Université de Bordeaux/RRI ORIGINS.

\newpage

\newpage

\newpage

\newpage


 \clearpage

\newpage
\bibliographystyle{aasjournal}
\bibliography{main}

\appendix
\renewcommand\thefigure{S.\arabic{figure}} \renewcommand\thetable{S.\arabic{table}}
\setcounter{figure}{0}
\setcounter{table}{0}
\subsection*{Thermodynamic integration and the calculation of the entropy}\label{Ap:TI}

MD simulations cannot directly estimate thermal quantities, such as free energy and entropy, of a non-ideal system \citep{frenkel_understanding_2002,leach_molecular_2001}. However, these quantities can be derived starting from the values of an ideal system, like an ideal gas.

We start with the free energy $F_0$ of an ideal gas system expressed as a function of its partition function as:

$$
F_0=-k_B T\times ln \frac{(N \times V)^N}{\Lambda^{3 N} N !}=-k_B T\left[N \ln \frac{V}{\Lambda^3}-\ln N !\right], 
$$
where $\Lambda$ is the Broglie wavelength, defined as:
\begin{equation}
    \\ \Lambda=\frac{h}{\sqrt{2 \pi m k_B T}},
\end{equation}
where k$_b$ is the Boltzmann constant, h the Planck constant, T the temperature, V the volume, and N the number of particles.

Then we express the free energy $F_1$ of the ab initio system starting from the ideal gas and adding a coupling term, $\Delta F_{0 \rightarrow 1}$, as in \cite{dorner_melting_2018}: 
\begin{equation}
    F_1=F_0+\Delta F_{0 \rightarrow 1}
\end{equation}

In practice, the coupling between the ab initio system and the ideal gas is realized using a linear mixing between the two descriptions of the dynamical states of the same system of particles. In this formalism, the forces, f, between the atoms can be evaluated as  a linear combination stretching form the pure ab initio system to the pure ideal gas system:

\begin{equation}
    f = \lambda f_1 + (1-\lambda) f_0
\end{equation}

$\lambda$ takes into account the part of the dynamics that comes from the ab initio system and (1-$\lambda$)  from the ideal gas. The coupling between the ab initio system and the ideal gas represents the integral along a thermodynamic path that couples these two systems:
\begin{equation}
    \Delta F_{0 \rightarrow 1}=\int^{\lambda=1}_{\lambda=0} \frac{\partial F}{\partial \lambda}
\end{equation}
which is equivalent to \citep{frenkel_understanding_2002,leach_molecular_2001}:
\begin{equation}
    \Delta F_{0 \rightarrow 1}=\int^{\lambda=1}_{\lambda=0}  d\lambda \Biggl \langle\frac{\partial U(\lambda)}{\partial \lambda} \Biggr \rangle_{\lambda} 
\end{equation}
and to \citep{dorner_melting_2018}:
\begin{equation}
    \Delta F_{0 \rightarrow 1}=\int^{\lambda=1}_{\lambda=0}  d\lambda \langle V(\lambda) \rangle_{\lambda} 
\end{equation}
where U is the internal energy, V is the potential energy, and the bracket denotes an average over a (well-thermalized and long-enough) simulation at constant $\lambda$.

$  \langle V(\lambda) \rangle_{\lambda} $ is estimated from a series of MD simulations. Once the free energy of the system is evaluated, the entropy becomes:
\begin{equation}
    S=\frac{U-F}{T}.
\end{equation}

\begin{table}[htpb]
    \centering
    \caption{Chemical composition of the three materials described in this study. CMc = chondrule composition, CMm = matrix composition, BSE = pyrolite/bulk silicate Earth composition.}
    \label{tab:compo}
    \begin{tabular}{ccccccccccc}
     & \multicolumn{6}{c}{Oxide (wt \%)} & Molar ratio & \multicolumn{3}{c}{Density (g.cm$^{-1}$)}\\
    & FeO  & SiO$_2$ & MgO   & Al$_2$O$_3$ & CaO  & Na$_2$O & (Mg+Fe)/(Si+Al) & $\rho_{0l}$ & $\rho_{0a}$ & $\rho_{0h}$ \\
        Chondrule (CMc) & 6.98 & 46.57   & 41.13 & 1.98        & 2.17 & 1.16    &  1.38 & 3.35 & 3.5 & 3.65 \\
        Matrix (CMm)    & 7.12 & 51.45   & 29.02 & 6.72        & 3.69 & 1.98    & 0.829  & 3.0 & 3.25 & 3.5\\
        Pyrolite (BSE)  & 8.94 & 44.68   & 37.23 & 4.75        & 3.48 & 0.93    & 1.25 & $\emptyset$ & 3.1 & $\emptyset$\\ 
    \end{tabular} 
\end{table}

\newpage

\begin{figure}[htbp]
    \centering
   \includegraphics[width=0.6\textwidth]{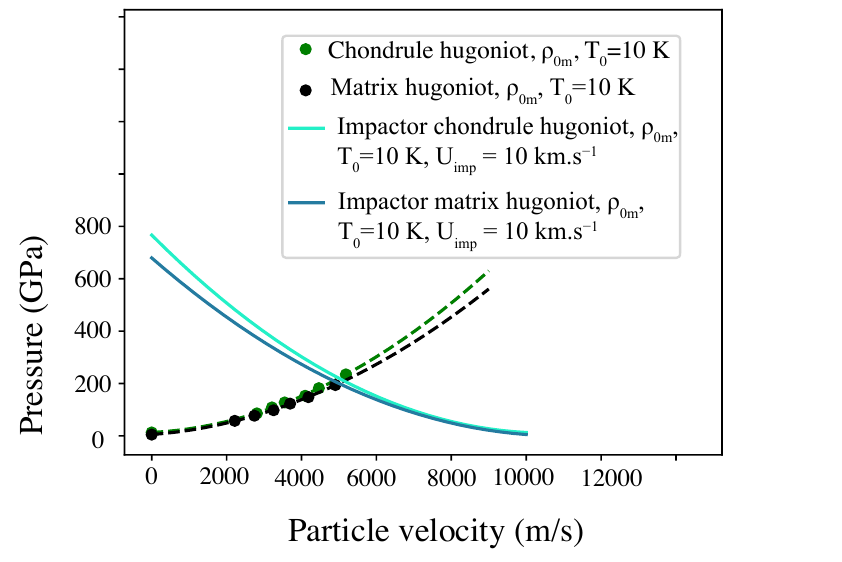}
\caption{Plot of P(u$_p$) and P(u$_{p_{imp}}$) for a CMm target and a CMc impactor traveling at 10 km/s.}
\label{fig:p(u)_plot}
\end{figure}

\newpage
\begin{figure}[htbp]
    \centering
    \includegraphics[width=1\linewidth]{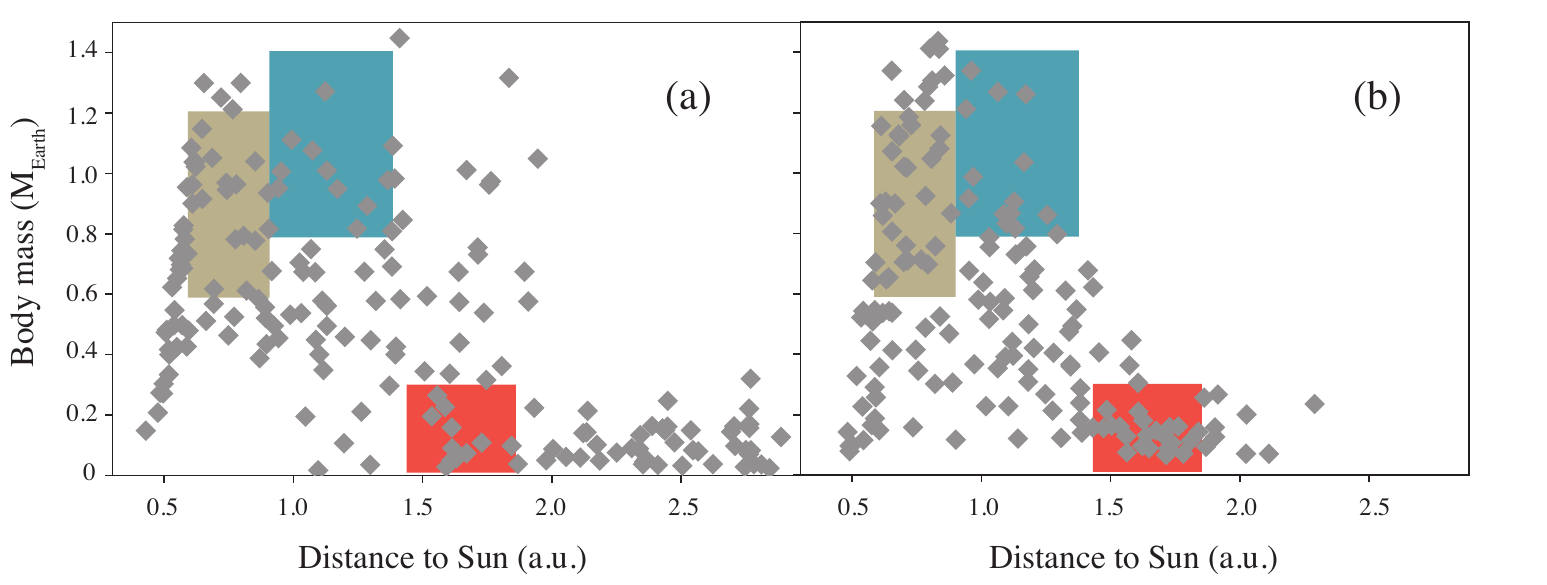}
    \caption{Remaining bodies at the end of the classical simulations (a) and annulus simulations (b). The yellow, blue, and red rectangles indicate, respectively, Venus, Earth, and Mars analog planets}
    \label{fig:nbody-planet}
\end{figure}

\newpage
\begin{figure}[htbp]
    \centering
    \includegraphics[width=0.9\textwidth]{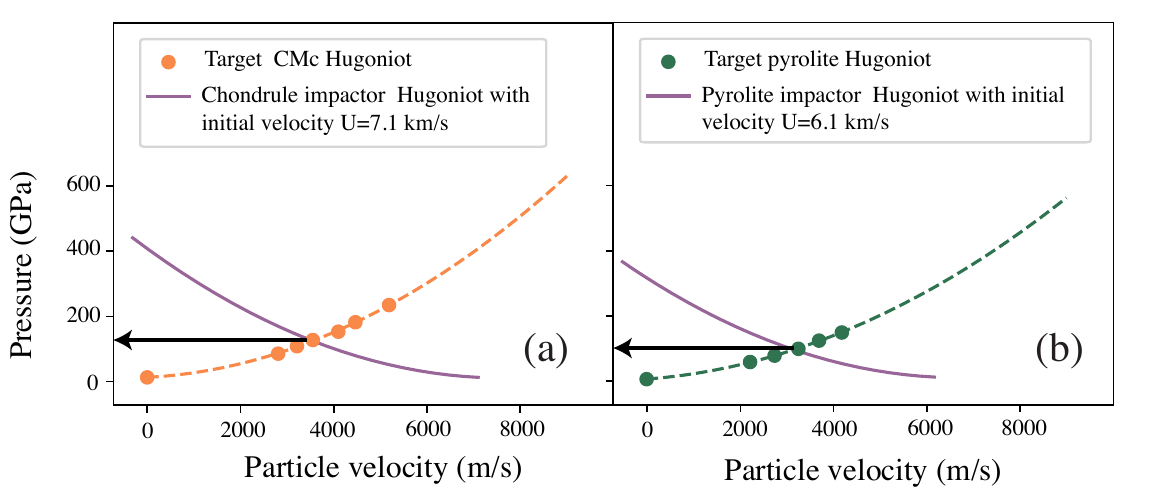}
    \caption{Determination of the peak impact pressure in the target in the planar shock collision approximation for chondrule (a) and pyrolite (b) material. The Hugoniot equations of the state of the two materials are plotted with opposite signs and cross at a given impact velocity $U_{mc}$ (respectively $U_{mp}$). The corresponding peak pressure, $P_{mc}$ (respectively $P_{mp}$), is read on the vertical axis (black horizontal arrow). Values are for impacts between two chondrule bodies and two pyrolitic bodies, respectively.}
    \label{fig:up-py}
\end{figure}

\newpage
\begin{figure}[htbp]
    \centering
    \includegraphics[width=0.9\textwidth]{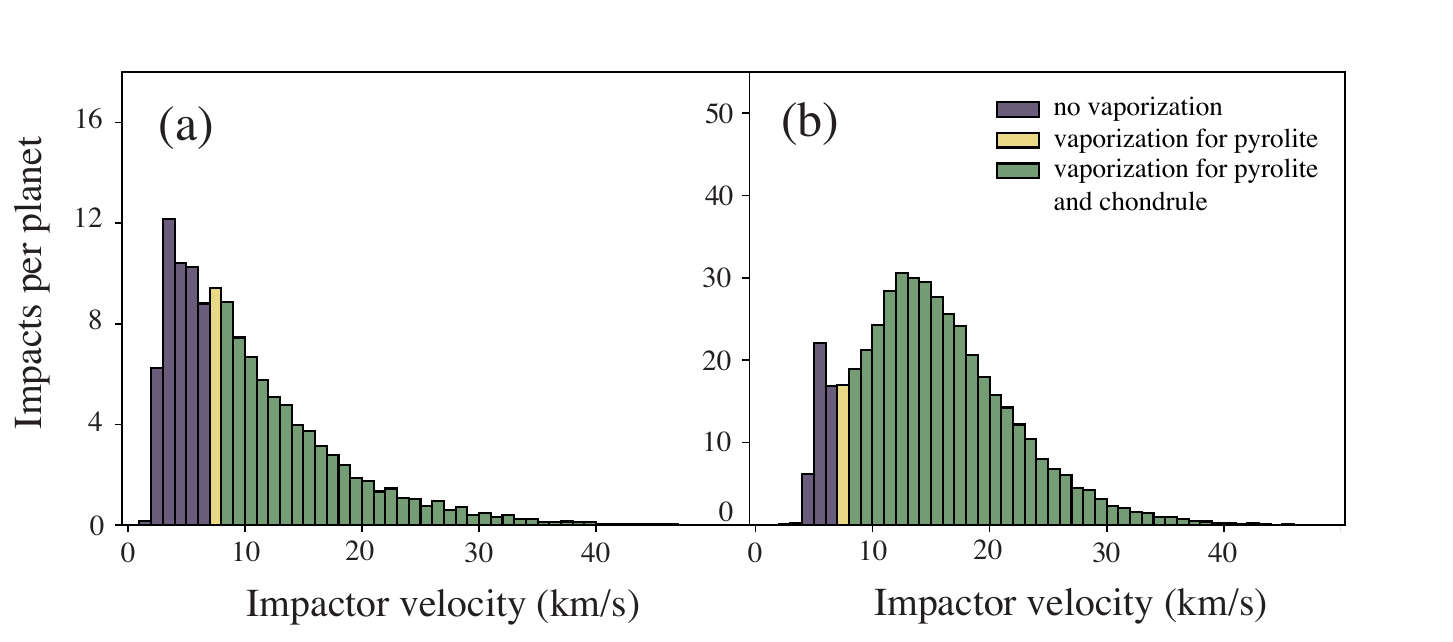}
    \caption{Proportions of impacts producing vaporization that occurred in N-body simulations with a classical starting configuration (a) and annulus starting configuration (b). The color represents the type of material that is vaporized.}
    \label{fig:histo-all}
\end{figure}

\newpage
\begin{figure}[htbp]
    \centering
    \includegraphics[width=0.95\textwidth]{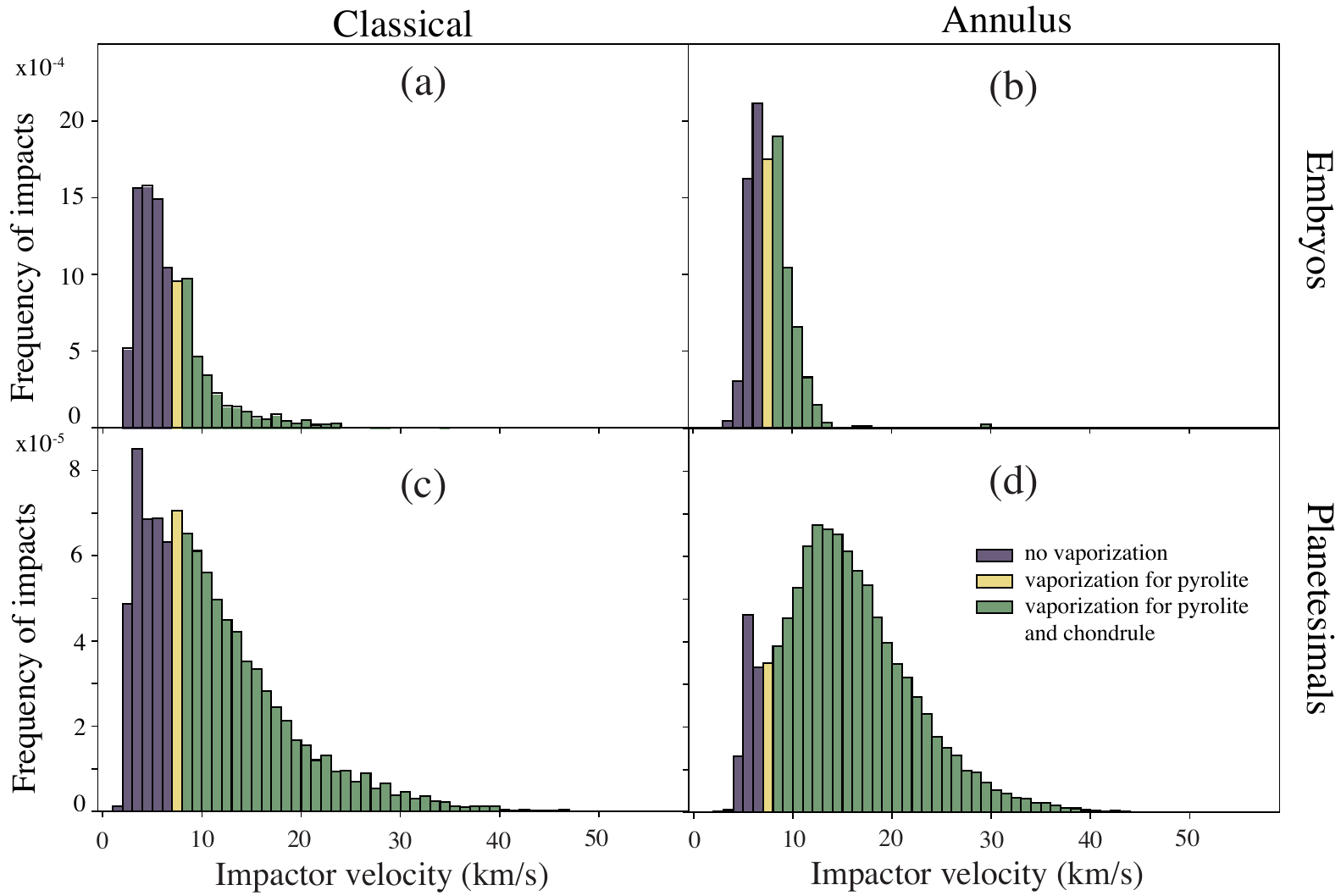}
    \caption{Proportions of impacts producing vaporization that occurred in N-body simulations in between two embryos (a and b) and in between an embryo and a planetesimal (c and d) with a classical starting configuration (a and c) and annulus starting configuration (b and d). Colors represent the type of material that is vaporized.}
    \label{fig:histo-embrypla}
\end{figure}

\begin{figure}[htbp]
    \centering
    \includegraphics[width=0.9\textwidth]{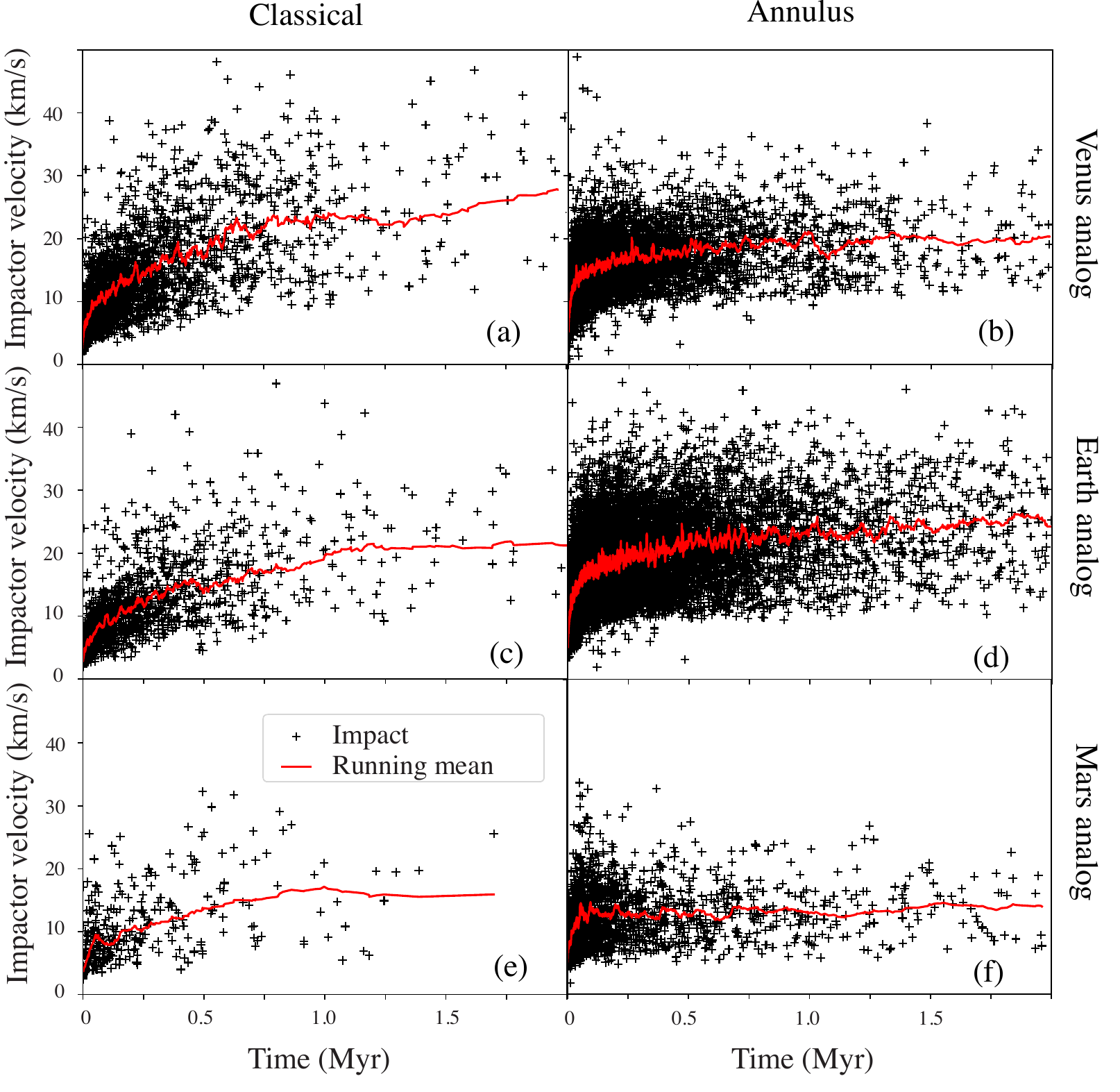}
    \caption{Relative velocities of impacts as a function of time for Venus (a and b), Earth (c and d), and Mars analogs (e and f) in N-body simulations with a classical starting configuration (a, c, and e) and annulus starting configuration (b, d, and f).}
    \label{fig:velocity}
\end{figure}

\end{document}